\documentclass[prl,twocolumn,showpacs,superscriptaddress]{revtex4}
\usepackage{epsfig}
\usepackage{color}

\begin{document}
\title{Percolative
phase transition on ferromagnetic insulator manganites:
uncorrelated to correlated polaron clusters}

\author{A.M.L. Lopes}
\email{armandina.lima.lopes@cern.ch}
\affiliation{CERN EP, CH 1211 Geneva 23, Switzerland.}
\affiliation{Departamento de F\'{\i}sica and CICECO, Universidade de Aveiro, 3810-193 Aveiro,
Portugal.}

\author{J.P. Ara\'ujo}
\affiliation{Departamento de F\'{\i}sica and IFIMUP,
Universidade do Porto, 4169-007 Porto, Portugal.}

\author{J.J. Ramasco}
\affiliation{Physics Department, Emory University, Atlanta GA 30322 USA.}

\author{E. Rita}
\affiliation{CERN EP, CH 1211 Geneva 23, Switzerland.}
\affiliation{ITN, E.N. 10, 2686-953 Sacav\'em, Portugal.}

\author{V.S. Amaral}
\affiliation{Departamento de F\'{\i}sica and CICECO, Universidade de Aveiro, 3810-193 Aveiro,
Portugal.}

\author{J.G. Correia}
\affiliation{CERN EP, CH 1211 Geneva 23, Switzerland.}
\affiliation{ITN, E.N. 10, 2686-953 Sacav\'em, Portugal.}

\author{R. Suryanarayanan}
\affiliation{Laboratoire de Physico-Chimie et de l'Etat Solide,
Universit\'e Paris--Sud, 91405 Orsay, France.}

\author{the ISOLDE Collaboration}
\affiliation{CERN EP, CH 1211 Geneva 23, Switzerland.}

\date{\today}

\begin{abstract}
In this work, we report an atomic scale study on the ferromagnetic
insulator manganite LaMnO$_{3.12}$ using $\gamma-\gamma$ PAC
spectroscopy. Data analysis reveals a nanoscopic transition from
an undistorted to a Jahn--Teller--distorted local environment upon
cooling. The percolation thresholds of the two local environments
enclose a macroscopic structural transition
(Rhombohedric--Orthorhombic). Two distinct regimes of
JT--distortions were found: a high temperature regime where
uncorrelated polaron clusters with severe distortions of the
Mn$^{3+}$O$_{6}$ octahedra survive up to $T \approx 800\, K$ and a
low temperature regime where correlated regions have a weaker
JT--distorted symmetry.
\end{abstract}

 %Pacs to be revised

\maketitle

Intense experimental and theoretical work has been devoted to
manganite systems due to their colossal magnetoresistance (CMR),
polaron dynamics and charge/orbital ordering phenomena. The
undoped manganites (AMnO$_3$ where A is a trivalent ion of La, Pr,
$\ldots$) typically show antiferromagnetic insulator behavior and
cooperative Jahn-Teller (JT) distortion of MnO$_6$ octahedra.
Oxygen excess or the presence of divalent ions at A--sites reduce
the static JT--distortion by the creation of Mn$^{4+}$ ions. This
effect favors the ferromagnetic interaction via dynamic electron
transfer between Mn$^{3+}$ and Mn$^{4+}$, the so called
double-exchange (DE) interaction \cite{Zener}. Although DE
interaction explains qualitatively the CMR, it does not fully
account for the large resistivity of the paramagnetic and
ferromagnetic insulator phases. Polaron formation must certainly
play an important role in this respect
\cite{Mill95,Bellinge,Zhao,Egami}. Polarons are formed due to the
strong electron-lattice coupling that leads to charge localization
via JT--distortions. Recently, the nature of such local
distortions, their dynamics and correlations have been addressed
by several authors
\cite{Kiryukhin1,Carron,Mannella,Kiryukhin2,Dagotto1}. In spite of
such an effort, several issues as the detailed structure of
polarons, the temperature evolution of polaron clusters or the
effect of such evolution on the average macroscopic lattice
structure still remain as open questions.

Local distortions and their dynamics can be studied by using
$\gamma-\gamma$ Perturbed Angular Correlation spectroscopy (PAC), a
nuclear hyperfine method specially effective to sample atomic scale
environments. PAC efficiency is $T$ independent, allowing to explore
a wide range of temperatures. To gain further insight on the
microscopic nature of polaronic distortions, their spatial
correlations and the role of polarons in ferromagnetic insulator
manganites (FMI), we have studied in detail the compound
LaMnO$_{3.12}$ using PAC technique. This compound is a prototypical
FMI manganite that undergoes a Rhombohedral ($R$)-Orthorhombic ($O$)
structural transition around room temperature, which provides us
with an ideal scenario to probe the evolution of local lattice
distortions through different average lattice symmetries. In
particular, we show that random distributed polaron clusters survive
in the undistorted $R$ phase up to temperatures as high as $766\,
K$. These distortions are as strong as those observed in the orbital
ordered LaMnO$_3$. Lowering $T$, the clusters continuously expand
until a microscopic transition takes place at $T_s \approx 170 \,
K$. Below the transition, the distortions are accommodated into a
weaker JT--distorted phase.

LaMnO$_{3+\Delta}$ polycrystalline samples ($\Delta = 0,\, 0.08$ and
$0.12$) were produced by the solid state reaction method. Powder
x-ray diffraction measurements show that the samples are chemically
homogeneous. In agreement with Refs. \cite{Ritter,Prado}, we find an
antiferromagnetic insulator ground state for the orthorhombic
JT--distorted $\Delta = 0$ compound ($T_N \approx 139\, K$), a
ferromagnetic insulator behavior for the weakly distorted $\Delta =
0.08$ sample ($T_c \approx 150\, K$) and a ferromagnetic insulator
state for the compound with $\Delta = 0.12$ ($T_c \approx 145\, K$).
This latter system presents as well an $O$-$R$ phase transition
around room temperature. As shown in Refs. \cite{Roosmalen,Ritter},
the oxygen excess $\Delta$ results in equivalent amounts of $La$ and
$Mn$ vacancies, with the fraction of $Mn^{4+}$ equal to $2 \Delta$.
$\gamma-\gamma$ PAC measurements were performed using a
high-efficiency 6-BaF$_2$ detector spectrometer \cite{Butz}. PAC
samples (one per measurement) were implanted at room temperature
with $^{111m}$Cd to a homogeneous low dose of $10^{12} \, cm^{-2}$
at $60 \, keV$ in the ISOLDE/CERN facility. Remaining point defects
created during implantation were eliminated by annealing at $700
\,^{\circ}C$ under O$_2$ controlled atmosphere for $20$ minutes. The
peak density of probing Cd only attains $1\,ppm$ of the La
concentration. Consequently, the implanted Cd atoms are simply
incorporated into La vacancies. The perovskite A (La) site is
specially appropriate to detect lattice distortions in the
surrounding MnO$_{6}$ octahedra since slight changes in the charge
geometry will significantly alter the EFG parameters.

\begin{figure}
\begin{center}
\label{fig1} \epsfysize=7cm \epsfbox{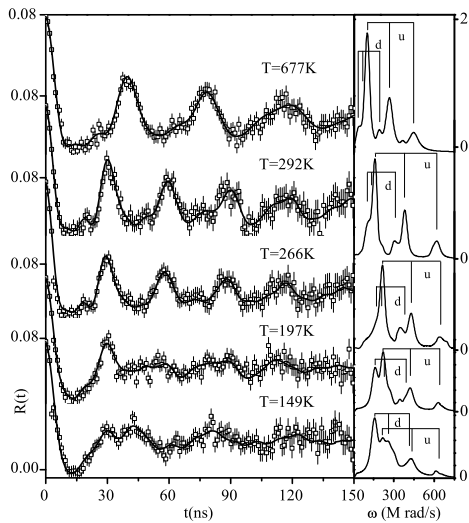} \caption{
Representative $R(t)$ experimental functions and the correspondent
fits for LaMnO$_{3.12}$. Corresponding Fourier transforms are
displayed on the right side.}
\end{center}
\end{figure}

The $^{111m}$Cd probes decay to $^{111}$Cd through an intermediate
state by the emission of two consecutive $\gamma$ rays. The half
life for the $^{111m}$Cd isomeric state is $T_{1/2} = 48\, min$,
while for the intermediate state is $T_{1/2} = 84\, ns$. The
angular correlation between the two $\gamma$ rays can be perturbed
by both the EFG and the Magnetic Hyperfine Field (MHF). These fields
respectively couple to the nuclear electric quadrupole
($Q$) and the magnetic
dipole ($\vec{\mu}$) moments of the intermediate nuclear state.
The Hamiltonian for such static interactions, in the proper
reference frame of the EFG tensor $V_{ij}$ with $|V_{zz}| \ge
|V_{yy}| \ge |V_{xx}|$, reads
\begin{equation}
\mathcal{H} = \frac{\hbar \,\omega_0 }{6} \left[ 3 \,I_z^2 - I\,
(I+1) + \frac{1}{2} \eta (I_+^2 + I_-^2) \right] +  \vec{\mu} \cdot
\vec{B}_{hf} ,
\end{equation}
where $\omega_0 = 3 \, e \, Q\,  V_{zz}/(2\, I\, (2\,I-1)\, \hbar)$
is the fundamental precession frequency, $I$ represents the nuclear
spin of the probe intermediate state ($I=5/2$ for $^{111}$Cd), $\eta
= (V_{xx}-V_{yy})/V_{zz}$ is the EFG asymmetry parameter and
$\vec{B}_{hf}$ is the magnetic hyperfine field \cite{schatz}. The
perturbation of the $\gamma-\gamma$ directional correlation is
described by the experimental $R(t)$ function, where $t$ is the time
spent by the nucleus in the $^{111}$Cd intermediate state. For a
hyperfine interaction, $R(t)$ may be expanded as $R(t) = \sum
A_{kk}\, G_{kk}(t)$ with $A_{kk}$ being the angular correlation
coefficients. The perturbation factor $G_{kk}(t)$ is the signature
of the fields interacting with the probes: MHF and a EFG in the
ferromagnetic phase and EFG alone for $T > T_c$. Below $T_c$, in the
presence of the two fields, we apply combined interaction theory to
obtain the MHF and EFG parameters. Above $T_c$, on the other hand,
$G_{kk}(t)$ may be expressed as \cite{schatz}
\begin{equation}
\label{Rt-stat} G_{kk}(t) =\ S_{k_0}+\sum_{n} S_{k_n} \,
\cos{(\omega_n \, t)} \, e^{- \omega_n \,\delta \, t}
\end{equation}
considering only pure electric quadrupole interactions.
The frequencies $\omega_n$ and amplitudes $S_{k_n}$
are determined by the $\mathcal{H}$ diagonalization. For spin $I=5/2$, three
frequencies are observable that are function of $\omega_0$ and
$\eta$ \cite{Butz2}. The exponential term in equation
(\ref{Rt-stat}) accounts for an attenuation of the $R(t)$ function
that appears in all spectra. This effect is due to randomly
distributed intrinsic vacancies and defects that produce a
Lorentzian distribution of static EFGs with central value
$\omega_0$ and relative width $\delta$. Independently, in
manganites, short-range charge diffusion coupled to lattice
distortions (polarons) can lead to EFG fluctuations. These
fluctuations contribute to further attenuate $R(t)$ when their
time scale is comparable to the life time of the PAC probe
intermediate state. When the characteristic fluctuation time
($\tau$) is shorter than the nuclear spin precession time ($2\,
\pi/\omega_0$), the $R(t)$ function can be satisfactorily
approximated by a single exponential damping term $e^{-\lambda \,
t}$ multiplying the static expression (\ref{Rt-stat}) with
$\lambda \propto \omega_0^2 \, \tau$ \cite{Baubry}.

Some experimental $R(t)$ curves are displayed in Fig. 1 for the compound
with $\Delta = 0.12$. We find in the
temperature range from $10\,K$ to $766\,K$ the coexistence of
three main local environments ($u, d, r$), {\it i.e.} three
fractions of probes ($f_u, f_d, f_r$) interacting with different
local EFG distributions.
\begin{figure}
\begin{center}
\epsfysize=7 cm \epsfbox{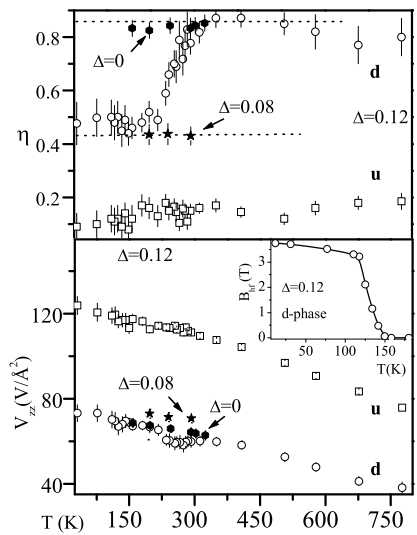} \caption{Asymmetry parameter
$\eta$ (top) and EFG principal component $V_{zz}$ (bottom) for
LaMnO$_{3.12}$ as a function of $T$. EFG parameters for
$\Delta=0.08$ and $\Delta=0$ are also showed. Inset: $T$
dependence of the MHF for the $d$ environment.}
\end{center}
\label{fig2}
\end{figure}
The environment $r$ is detected by a low residual fraction of the Cd
probes ($5\%$), which is temperature independent. Its EFG parameters
are approximately $V_{zz}^r \approx 102 \,$ V/\AA$^2$ and $\eta_r
\approx 0.9$ at room temperature. These values suggest a very asymmetric local
charge distribution. The origin of this environment might be related
to probes located at the vicinity of Mn/La vacancies
 and/or other defects. Actually, assuming that the positions of the
 vacancies are not correlated, the
probability that a Cd siting in a La vacancy has in its surrounding
a Mn or next shell La vacancies is roughly $2\%$.

In Fig.2, the temperature dependence of the EFG asymmetry parameter
$\eta$ (top) and principal component $V_{zz}$ (bottom) for the $u$
and $d$ environments is displayed. For comparison, the EFG
parameters found in $\Delta=0.08$ and $\Delta=0$ samples are also
included in the same figure. The $u$ environment that is dominant at
high $T$ shows an almost axially symmetric EFG ($\eta_u \approx 0
$). This value characterizes an EFG with an axis of threefold or
higher rotational symmetry, which is compatible with the Rhombohedral
lattice structure observed at high temperature. The MnO$_6$
octahedra in the $R$ structure are constrained by symmetry to be
JT--undistorted (equal Mn-O bond lengths), thus we will name this
local environment {\it undistorted}. In contrast, the $d$ ({\it
distorted}) environment is characterized by a weaker $V_{zz}$
\cite{note} and highly asymmetric EFG ($\eta_d > 0.45$). At high
$T$, the values of $\eta_d$ and $V_{zz}^d$ coincide with the ones
observed for the undoped fully JT--distorted Orthorhombic system,
$\Delta=0$, (full circles in Fig. 2). Consequently, at high
temperatures, the $d$ local environment must be characterized by a
distortion involving several (minimum eight) Mn$^{3+}$O$_{6}$
octahedra similar to the collective JT-distorted lattice of the
orbital ordered LaMnO$_3$ \cite{lmo3}. Lowering $T$ below $300 \,
K$, the asymmetry parameter $\eta_d$ decreases stabilizing at a
value close to that observed for the $\Delta=0.08$ sample (solid
stars in Fig. 2). This behavior suggests that the JT--distortions
are weakening till they reach a similar degree as in the
$\Delta=0.08$ sample. The EFG principal components $V_{zz}^u$ and
$V_{zz}^d$ slightly increase with decreasing temperature. This is a
typical feature of perovskite and related systems \cite{Dogra}.
Below $T_c\approx 145\, K$ both $d$ and $u$ local environments
experience increasing magnetic hyperfine fields upon decreasing
temperature (inset of Fig. 2), presenting at $10 K$ values of
$B_{hf}^d = 3.8(2)T$ and $B_{hf}^u = 4.0(3)T$ compatible with a full
ferromagnetic environment of the surrounding Mn ions
\cite{Gubkin,Allodi}.

%%(attributed to the variation of the atomic lattice vibrations with
%%temperature)

\begin{figure}
\begin{center}
\epsfysize=5.5cm \epsfbox{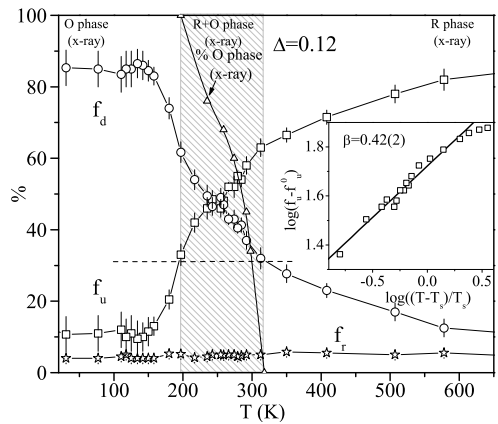} \caption{Temperature dependence
of the probe volume fractions $f_u$, $f_d$ and $f_r$. Triangles:
orthorhombic phase percentage from x-ray diffraction. The shadowed
region is limited by the temperatures where the percolation
thresholds occur. Inset: log-log plot of $(f_u-f^0_u)$ vs
$(T-T_s$).}
\end{center}
\label{fig3}
\end{figure}

Further insight in the behavior of $d$ and $u$ environments may be
achieved by studying the $T$ dependence of the volume
fractions $f_u$ and $f_d$. As may be seen in Fig. 3, the $u$
environment is dominant at very high temperatures ($f_u \approx
86\%$ at $T = 766\, K$), though $d$ regions survive up to that $T$
($f_d \approx 9\%$). This confirms the high stability of the
inhomogeneous phase-segregated state. Our data, at high T, are
compatible with a scenario where random distributed JT--distorted
nanoclusters are embedded in a undistorted matrix as predicted by
\cite{Dagotto}. At very low $T$, the fraction of $u$ environment
reaches a remanent value ($f_u^o \approx 10 \%$), which is typically
observed in CMR manganites \cite{savosta} and is a signature of the
ferromagnetic-metallic (FMM) and FMI phase coexistence. When the
temperature changes, $f_u$ (symmetrically $f_d$) suffers a smooth
variation leading
from an undistorted to a JT--distorted dominant microscopic
environment. If we assume
that this variation is a continuous phase transition, the order
parameter would be $f_u-f_u^o$ and must follow a power law behavior
$f_u-f_u^o \sim (T-T_s)^\beta$ when the critical temperature $T_s$
is approached from above. To check this possibility, we display $f_u$
in a log-log plot in Fig. 3 (inset). The data
adjust pretty well to a power law with $T_s \approx 170 \pm 10 K$
(relatively close to $T_c$) and $\beta \approx 0.42 \pm 0.02$.
Associated to the transition, there must also exist a correlation
length, correlation of the $d$ spatial distribution, that must
diverge at $T_s$. As may be seen in Fig. 2, when $T$ decreases
$\eta_d$ starts to fall as the $d$ component percolates (at $f_d
\approx 31.16 \%$ \cite{stauffer}) and only stabilizes around $T_s$.
Macroscopically, on the other hand, x--rays measurements detect a
structural transition ($R$--$O$) that lies exactly between the
temperatures corresponding to the percolation thresholds of the two main
nanoscopic components. These are precisely the temperatures in which
the minority invading cluster suffers a sudden size divergence becoming
macroscopically observable.

\begin{figure}
\begin{center}
\epsfysize=5.5cm \epsfbox{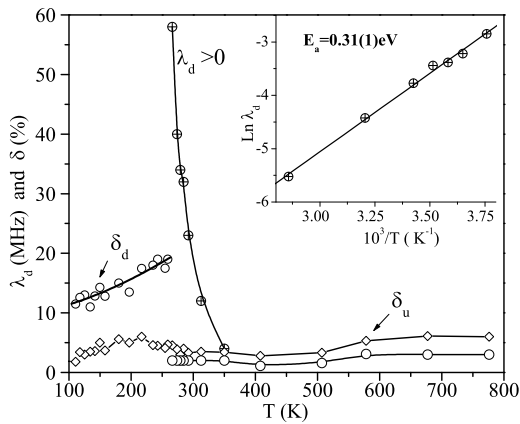} \caption{Temperature
dependence of static $\delta_d$ ($\circ$)  and dynamic $\lambda_d$
($\oplus$) attenuation parameters to the R(t) function for $d$
environment. Static attenuation  $\delta_u$ ($\diamond$) for $u$
environment. Inset: Arrhenius plot of $\lambda_d$ to estimate the
activation energy.}
\end{center}
\label{fig4}
\end{figure}

The temperature dependence of the attenuation of $R(t)$ provides
additional information about the dynamics of the $u$ and $d$
environments. A complete sketch of the dynamic and static
attenuation for $R(t)$ in both environments is depicted in Fig. 4.
The best fit to the R(t) spectra discards the presence of time
dependent interactions for the $u$ environment ($\delta_u \approx
4\%$ independently of $T$ and $\lambda_u = 0$). Thus, in all
temperature range, the charge transfer between Mn$^{3+}$ and
Mn$^{4+}$ (activated hopping) in this environment should occur with
a frequency higher than we can probe. For the $d$ environment, on
the other side, the best fits were obtained admitting a fluctuating
EFG ($\lambda_d \ne 0$ and $\delta_d = 2\%$) in the temperature
region spanning from $T= 266 \, K$ to $T = 350 \, K$. Notice that
these time dependent effects cannot be attributed to Cd/O and/or
defects diffusion since they would be detected in both fractions.
The temperature dependence of the dynamic attenuation parameter,
$\lambda_d$, allows us to estimate an activation energy $E_a$. This
energy is obtained from $\lambda_d = \lambda_\infty \, e^{E_a/kT}$,
and was found to be $E_a \approx 0.31 \, eV$ (see inset Fig. 4),
close to the polaron binding energy reported in the literature for
low doped manganites \cite{Goodenough,Weibe}. We identify such EFG
fluctuations with polaron diffusion related to charge (hole)
transport. The EFG fluctuation time ($\tau$) can be estimated from
the maximum of $\lambda_d (T)$ \cite{Winkler}. Considering that a
carrier (hole) can hop to any of the $8$ octahedra around a La site
(8 possible EFG states), we find $\tau = 0.5\, \mu s$ at $T = 266\,
K$ corresponding to ultra--slow polaron diffusion. Similar polaron
residence times have been recently reported in \cite{Allodi},
although the $E_a$ measured there was smaller possibly due to the
intense magnetic field ($7\, T$) needed to perform NMR measurements.
The competition of the distinct dynamics of the $u$ (fast hopping)
and $d$ (related to polaronic conduction) environments is
responsible for the macroscopic ferromagnetic insulator behavior
observed in these systems \cite{savosta}. Below $T_c$, both local
environments become ferromagnetic and a phase coexistence between
metallic ($u$) and insulator ($d$) regions exists. However, the
majority fraction ($d$) is characterized by ultra--slow diffusion of
charge carriers imposing an overall insulator behavior.

In conclusion, we report an extensive study on the ferromagnetic
insulator manganite LaMnO$_{3.12}$ using $\gamma-\gamma$ PAC
spectroscopy. We analyze in detail the evolution and stability of
polaron clusters in an extremely wide range of $T$ that includes a
structural transition between $R$ and $O$ phases. PAC measurements
reveal a continuous transition between two different dominant local
atomic environments (one JT--distorted ($d$) and another undistorted
($u$)). Information is also obtained on the local structure, the
dynamics and the correlations of these two environments. The
macroscopic transition arises as a consequence of the microscopic
changes, since it occurs between the percolation thresholds of the
two local components. The $d$ environment survives up to very high
$T$ where it can be identified with uncorrelated polaron clusters.
The correlation of $d$ clusters increases for $T$ below the $d$
percolation threshold diverging at $T_s$. These results provide
further insight in the understanding of the nature/evolution of
polaronic distortions and FMM-FMI phase competition responsible for
the insulator behavior of these systems.

The authors gratefully thank W. Troeger, U. Wahl, J.M. L\'opez, R.
Valiente, J. Vieira and R. Catherall for fruitful discussions. This
work was funded by the Portuguese Research Foundation (FCT), FEDER
(projects POCTI/FN/FNU/50183/03, PDCT/FP/FNU/50145/2003), and EU
(Large Scale Facility contract HPRI-CT-1999-00018). J.J.R. received
partial funding from the NSF under grant 0312510. A.M.L.L.
and E.R. acknowledge their grants to FCT.


\begin{thebibliography}{99}
\bibitem{note} Note that the fact that $V_{zz}^d < V_{zz}^u$ alone
is not proof enough of a higher symmetry. It has been experimentally
observed \cite{Dogra} a sudden decrease in the $V_{zz}$ and increase
in $\eta$ related to a change from a $R$ to a $O$ structure in
stoichiometric LaCrO$_3$ and LaFeO$_3$ perovskites.
\bibitem{lmo3} When the Sternheimer corrections are taken into account, our values
for the EFG parameters agree with the results reported in the
literature on the undoped compound using NMR \cite{Allodi} and from
ab-initio FLAPW theoretical calculations \cite{Ravi}
\end{thebibliography}
\end{document}